\documentclass[sts]{imsart}

\RequirePackage{amsthm,amsmath,amsfonts,amssymb}
\RequirePackage[numbers,sort&compress]{natbib}
\RequirePackage[colorlinks,citecolor=blue,urlcolor=blue]{hyperref}
\RequirePackage{graphicx, subcaption}
\usepackage{centernot, mathtools}
\usepackage{lineno}

\startlocaldefs
\theoremstyle{plain}

\newtheorem{theorem}{Theorem}[section]

\theoremstyle{definition}

\newcommand{\ind}{\mathrel{\perp\!\!\!\perp}}

\newcommand{\notind}{\mathrel{\centernot{\ind}}}

\newcommand{\cond}{\mathrel{\,|\,}}

\newcommand{\edge}[1][.7em]{\mathrel{\rule[.5ex]{#1}{.4pt}}}

\endlocaldefs

\begin{document}

\begin{frontmatter}
\title{Discovering causal relationships between time series with spatial structure}

\runtitle{Causal discovery for spatial time series}

\begin{aug}
\author[A]{\fnms{Rebecca F.}~\snm{Supple}\thanks{[\textbf{Corresponding author}]}\ead[label=e1]{rbs9@st-andrews.ac.uk}\orcid{0000-0002-9752-0656}},
\author[B]{\fnms{Hannah}~\snm{Worthington}}
\and
\author[C]{\fnms{Ben}~\snm{Swallow}}

\address[A]{Rebecca F. Supple is PhD Student, Centre for Research into Ecological and Environmental Modelling, School of Mathematics and Statistics, University of St Andrews, St Andrews, United Kingdom\printead[presep={\ }]{e1}.}

\address[B]{Hannah Worthington is Senior Lecturer, Centre for Research into Ecological and Environmental Modelling, School of Mathematics and Statistics, University of St Andrews, St Andrews, United Kingdom.}

\address[C]{Ben Swallow is Lecturer, Centre for Research into Ecological and Environmental Modelling, School of Mathematics and Statistics, University of St Andrews, St Andrews, United Kingdom.}

\end{aug}

\begin{abstract}
    Causal discovery is the subfield of causal inference concerned with estimating the structure of cause-and-effect relationships in a system of interrelated variables, as opposed to quantifying the strength or describing the form of causal effects. As interest in causal discovery builds in fields such as ecology, public health, and environmental sciences, where data are regularly collected with spatial and temporal structures, approaches must evolve to manage autocorrelation and complex confounding. As it stands, the few proposed causal discovery algorithms for spatiotemporal data require summarizing across locations, ignore spatial autocorrelation, and/or scale poorly to high dimensions. Here, we introduce our framework to extend time-series causal discovery to systems with spatial structure. Building upon work on causal discovery across contexts and methods for handling spatial confounding in causal effect estimation \cite{Gilbert2024CISpatConf, Gunther2023JointPCMCI}, we prove the soundness and completeness of our extension in the large sample limit. We close by discussing the feasibility of the framework, outlining remaining gaps in the literature, and identifying directions for future research.
\end{abstract}


\end{frontmatter}

Identification of cause-and-effect relationships from observational data, termed \textit{causal discovery}, has always been at the center of scientific research \cite{Pearl2009Causality}. In causal discovery, a system $\mathcal{X}$ of potentially interrelated variables of which we observe $\mathbf{X} = \{X_i\}_{i \in \mathcal{I}}$, $\mathcal{I} = \{1, ..., p\}$, is analyzed to determine whether each pair of variables $X_i, X_j \in \mathbf{X}, i \neq j$ has a \textit{causal} relationship whereby an intervention in/manipulation of $X_i$ leads to change in $X_j$. A causal discovery algorithm outputs an estimate of where (if any) causal relationships exist in the system (e.g. Figure~\ref{Graph Example}), providing key insight for causal model building, effect estimation, and/or counterfactual inference \cite{Pearl2009Causality}. 

Statistical methods with this explicit goal have only emerged in the last three decades and are still catching up to the complexity of practically available data \cite{Cinelli2025Challenges, Glymour2019CDReview}. Spatiotemporal data pose particular difficulties for causal discovery methods due to autocorrelation between observations and latent confounders that may mask or distort relationships \cite{Ninad2025VectorValuedCD}. In the following sections, we introduce causal discovery and motivate its extension to spatiotemporal systems. We then show how existing methods are insufficient to address spatial structures out-of-the-box, and show how results concerning unobserved spatial confounding in causal effect estimation allow us to extend time series causal discovery to systems with spatial structure. We close by addressing the strengths and limitations of this extension and identifying promising areas of future research.

\section{Causal discovery}\label{CD Intro}

\begin{figure}
    \centering
    \includegraphics[width=0.5\linewidth]{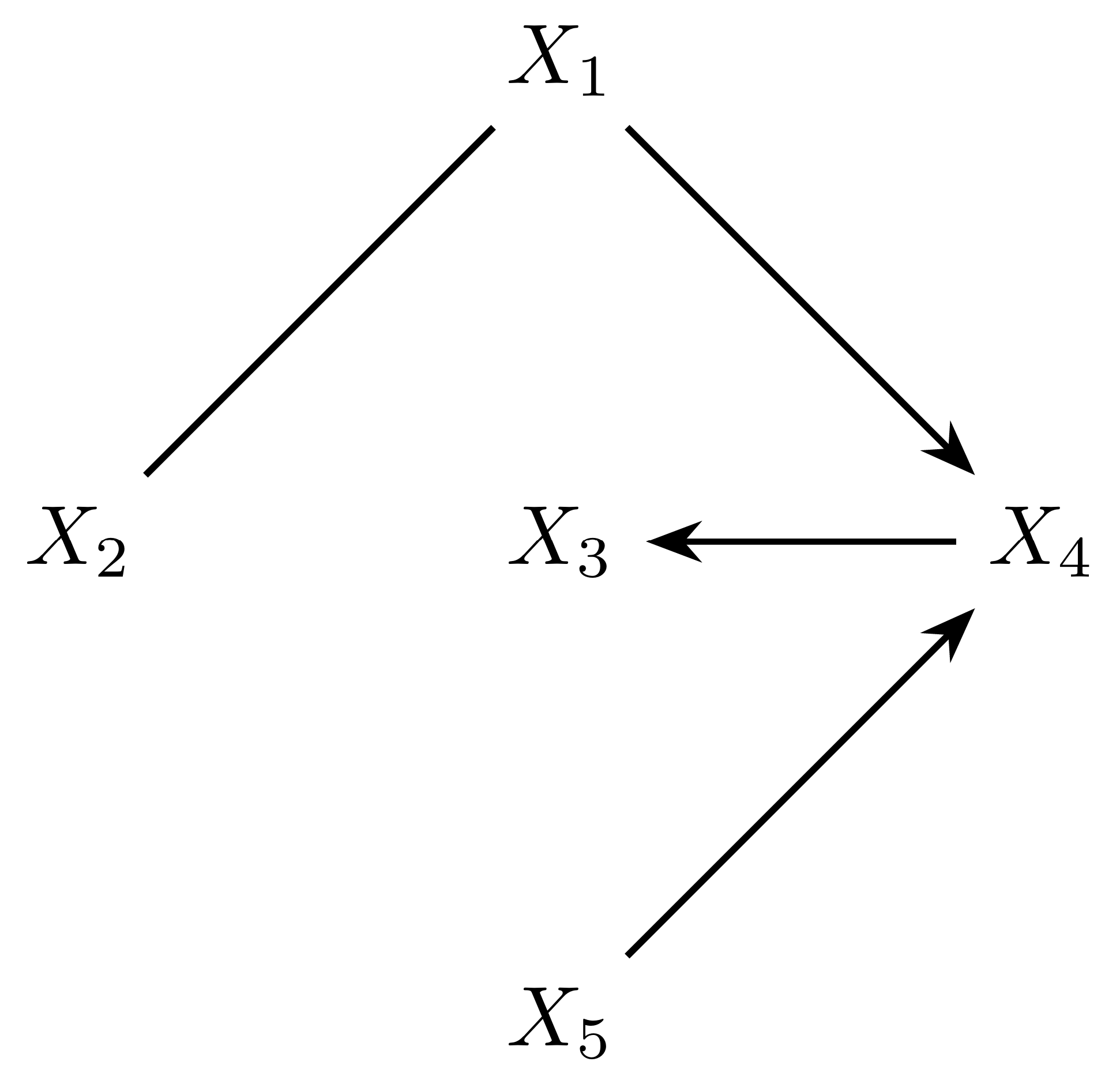}
    \caption{An example causal graph estimated via constraint-based causal discovery for an observed system $\mathbf{X} = \{X_1, X_2, X_3, X_4, X_5\}$. An edge directed $X_i \rightarrow X_j$ indicates that $X_i$ causes $X_j$, or, equivalently, $X_i \in \mathbf{pa}(X_j)$. An undirected edge $X_i \edge X_j$ indicates a causal connection where the direction is unresolvable (i.e.~$X_i$ may cause $X_j$ or vice versa).}
    \label{Graph Example}
\end{figure}

The graphical causal approach proposes that systems of variables $\mathbf{X}$ with relationships defined by the data-generating structural causal model (SCM) $\mathcal{M}$ can be represented by graphs \cite{Pearl2009Causality}. In such graphs, nodes corresponding to each variable $X_i \in \mathbf{X}$ are connected to nodes corresponding to their immediate causes, or parents $\mathbf{pa}(X_i) \subset \mathbf{X}$ by directed edges (arrows) without forming feedback loops (Figure~\ref{Graph Example}) \cite{Verma1990DAGs, Pearl1998SEMs}. Causal graph structures are often defined by subject expert consensus or as a representation of a specific hypothesis \cite{Petersen2023DatavTheoryLongitudinalCD}. In either case, some confirmation bias may be present as only hypotheses previously considered in the literature and/or by the modeller can be put forth. There are also instances where no good hypotheses exist \textit{a priori} to describe a system and exploratory analysis must be undertaken, motivating estimation of a causal graph from observational data.

There are three principal approaches to causal discovery. Score-based causal discovery methods estimate a best-fit graph by minimizing a causally-relevant loss function across the space of possible graphs \cite{Chickering2002GreedySearch, Janzing2010CIMarkov}. Functional methods exploit asymmetries in non-Gaussian noise to discover causal connections and directions \cite{Shimizu2006LiNGAM}. Constraint-based methods iterate through all pairs of variables, testing their independence conditional on other variables in the system and estimate cause-and-effect relationships by eliminating structures that are inconsistent with conditional independencies and/or do not meet assumptions \cite{Spirtes1993DiscoveryReview, Pearl2009Causality}. We focus on constraint-based methods as they do not impose any inherent parametric or functional assumptions and thus maintain flexibility in the complex spatiotemporal structures considered further here \cite{Glymour2019CDReview, Cinelli2025Challenges}.

\subsection{Constraint-based causal discovery}

The Peter-Clark (PC) algorithm was proposed in 1991 as the first asymptotically correct constraint-based causal discovery algorithm that could handle more than a few independently and identically distributed (iid) variables \cite{Spirtes1991PC}. In the decades since, the logic of the PC algorithm has been a popular starting point for extensions \cite{Ramsey2006Faithfulness, Runge2019PCMCI, Gunther2023JointPCMCI, Spirtes1999FCI, Saggioro2020RegimeCD}. 

For PC and PC-derived algorithms, the goal is to estimate the placement and direction of edges in the causal graph of a system of variables. Given a system with $n$ iid observations of $p$ variables $\mathbf{X} = \{X_1,..., X_p\}$, the algorithm first supposes a fully connected graph with undirected edges between all pairs $X_i, X_j \in \mathbf{X}, i \neq j$. It then searches for separating sets $\mathbf{Z}_{ij} \subseteq \mathbf{X}\setminus\{X_i, X_j\}$ such that $X_i \ind X_j \cond \mathbf{Z}_{ij}$. Where (conditional) independence is found, or, equivalently, when such a set $\mathbf{Z}_{ij}$ exists, the variables $X_i, X_j$ must not be causally connected and the edge between them is deleted. 

Once the placement of edges has been determined, they are directed according to the following constraints. Where dependence between otherwise (conditionally) independent variables $X_i$ and $X_j$ is introduced by conditioning on a mutually adjacent variable $X_a$, a \textit{collider} or common cause is indicated and edges are directed $X_i \rightarrow X_a \leftarrow X_j$. For example, $X_1 \rightarrow X_4 \leftarrow X_5$ in Figure~\ref{Graph Example} would be indicated by the relationships $X_1 \ind X_5; X_1 \notind X_5 \cond X_4$. Once colliders have been determined, remaining undirected edges are directed to avoid inducing feedback loops/cycles and so as not to create false colliders. When direction is not resolvable according to the constraints edges remain undirected (e.g.~$X_2 \edge X_1$ in Figure~\ref{Graph Example} \cite{Pearl2009Causality}).

The PC algorithm assumes the following \cite{Spirtes1991PC, Pearl2009Causality}: 
\begin{itemize}
    \item \textit{Causal Markov condition}: Non-adjacency in the underlying causal graph implies that two variables are independent, conditional on their parents;
    \item \textit{Faithfulness}: Conditional independence corresponds to non-adjacency in the underlying causal graph;
    \item \textit{Sufficiency}: All relevant variables are observed; and
    \item \textit{Stability}: Relationships are consistent across observations.
\end{itemize}
Various extensions relax the faithfulness \cite{Ramsey2006Faithfulness}, sufficiency \cite{Spirtes1999FCI}, and/or stability \cite{Gunther2023JointPCMCI, Saggioro2020RegimeCD} assumptions.

\section{Causal discovery for spatiotemporal data}

Methods for causal discovery described so far assume all variables are iid, and that there are no confounding variables missing from the system. Data distributed across space and time are often autocorrelated, however, and may depend on unobserved time-lagged and spatial confounders \cite{Gerhardus2020LatentPCMCI, Gilbert2024CISpatConf}. 

These obstacles can be approached differently according to the goal or overarching question being asked. We introduce here an organizing principle to distinguish methods by the research question they seek to answer:
\begin{itemize}
    \item Latent pattern: \textit{how do changes in (a) variable(s) in this location affect (a) variable(s) in other locations across time?} \\
    \item Latent mechanism: \textit{what is the underlying structure of causal relationships between variables whose observations are distributed and autocorrelated across space and time?}
\end{itemize}
We illustrate these two approaches in Figure~\ref{fig:Illustrations}.

\subsection{Latent pattern causal discovery}

\begin{figure}
    \centering
    \begin{subfigure}{0.47\textwidth}
        \includegraphics[width=\textwidth]{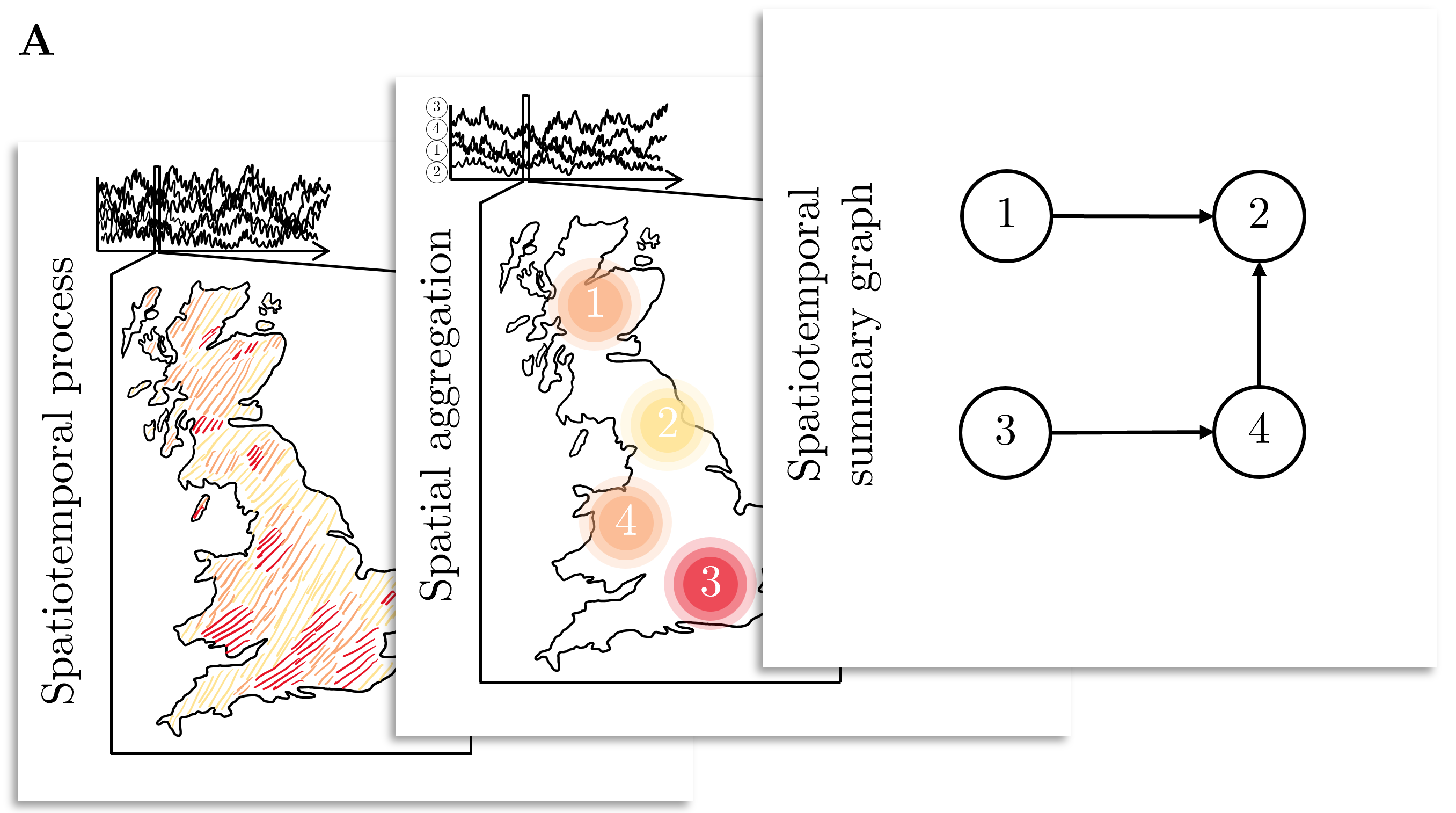}
        \label{fig:Latent Pattern}
    \end{subfigure}
    \begin{subfigure}{0.47\textwidth}
        \includegraphics[width=\textwidth]{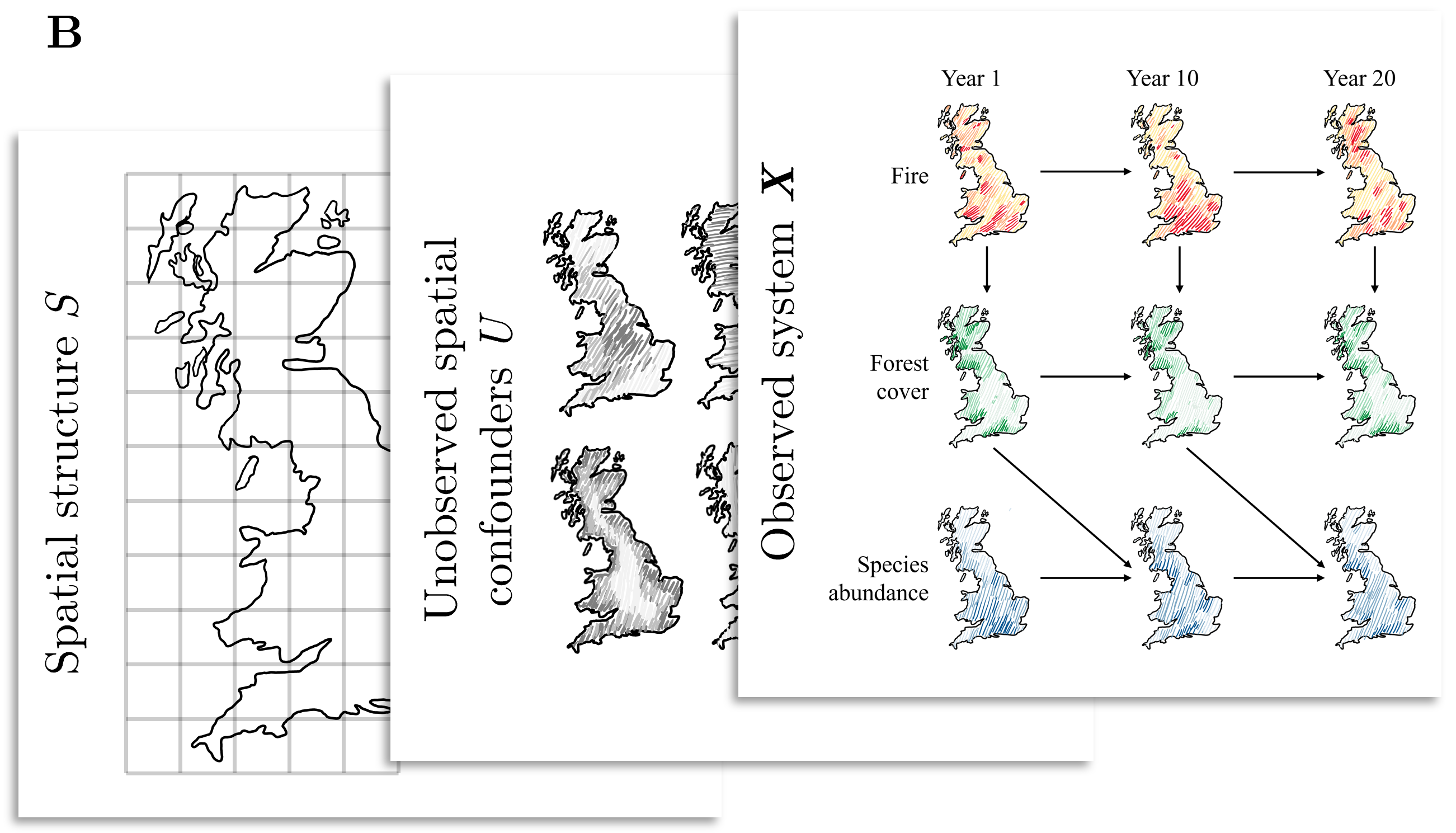}
        \label{fig:Latent Mechanism}
    \end{subfigure}
    \caption{Illustrations of systems relevant to (A) latent pattern causal discovery and (B) latent mechanism causal discovery. \\
    A) Latent pattern causal discovery may be concerned with estimating a graph of causal influence of different locations on each other across time. In this illustration, a spatiotemporal process (left) is summarized into four spatially uncorrelated time series (middle). A graph is then estimated from those time series to summarize the relationships between locations (right). \\
    B) Latent mechanism causal discovery is concerned with estimating the graph of causal relationships between variables given observations have temporal and spatial structure. This illustration mirrors the SCM described in Equation~\ref{eq:Latent Mech SCM}. A set of unobserved spatial confounders $U$ (middle) are defined as fixed, measurable functions of spatial structure $S$ (left). The observed system variables $\mathbf{X}$ (right) are functions of their observed parents as well as unobserved spatial confounders $U$. Latent mechanism causal discovery seeks a graph similar to that on the right panel summarizing causal relationships between observed spatial time series.}
    \label{fig:Illustrations}
\end{figure}

The current literature on causal discovery for spatiotemporal data tends to focus on estimating patterns of influence within and/or between a handful of variables on a large spatiotemporal scale \cite{Wang2025LatentGraphsFromSTData, Sheth2022STCDHydro, Sheth2023STCDStreamflow, Lippe2025STCausalRepLearning, Nichol2025STCDEarthScienceStencil}. Spatial autocorrelation is often ignored according to domain knowledge \cite{Sheth2022STCDHydro, Sheth2023STCDStreamflow} or removed by estimating effectively independent spatial factors that aggregate nearby points (Figure~\ref{fig:Illustrations}A) \cite{Wang2025LatentGraphsFromSTData, Lippe2025STCausalRepLearning, Oprescu2025STCIwTimeVaryingConf, Runge2015STCDReview, Tibau2022ClimateBenchmark}. While some methods allow for relationships to vary continuously with time \cite{Mameche2025NonStatTSCD}, others look at temporal snapshots \cite{Nichol2025STCDEarthScienceStencil}. The estimated causal patterns summarize observed phenomena and can, for example, inform extrapolative forecasting models.

Since these methods usually focus on changes across space and time for a single variable, latent confounders are often not explicitly addressed, only considered temporally, or assumed exogenous \cite{Nichol2025STCDEarthScienceStencil, Rabel2025LatentContext}. 

\subsection{Latent mechanism causal discovery}

A relatively understudied goal of causal discovery for spatiotemporal data is inferring the underlying structures of causal relationships between variables whose observations are distributed and autocorrelated across space and time \cite{Cinelli2025Challenges, Ninad2025VectorValuedCD}. Latent mechanism causal discovery seeks to identify the graph over the observed variables $\mathbf{X}$, whose relationships are stationary and do not vary across space and time, but whose distributions are partially determined by spatially-structured unmeasured confounders and potentially time-lagged parents. Latent mechanism methods may be more relevant for fields like ecology, economics, and public health, where questions about relationships between variables have direct policy and management implications and where observational data is often highly structured \cite{Petersen2021LongitudinalCD}. 

Here, we illustrate one potential SCM that could underpin latent mechanism causal discovery (Figure~\ref{fig:Illustrations}B). Let $\mathcal{X}$ be a system comprising space- and time-dependent variables $\mathbf{X}_t^{(s)} = \{X_{i,t}^{(s)}\}_{i \in \mathcal{I}}$ at times $t \in \{1,...,T\}$ and locations $s \in S$ as well as latent spatial confounders $\mathbf{U}_{t}^{(s)}$, such that, for each $i$ in the set of indices $\mathcal{I}$:
\begin{equation}\label{eq:Latent Mech SCM}
    \begin{split}
        & X_{i,t}^{(s)} := f(\mathbf{pa}(X_{i,t}^{(s)}), \mathbf{U}_{t}^{(s)}, \epsilon_{i,t}^{(s)}),\\
        & \mathbf{U}_{t}^{(s)} := \mathbf{h}(S)
    \end{split}
\end{equation}
for a set of fixed, measurable functions $\mathbf{h}$, and where all exogenous noise terms $\epsilon_{i,t}^{(s)}$ are jointly independent and identically distributed with respect to both time and space \cite{Gilbert2024CISpatConf}. Readers can compare this approach with latent pattern in more detail by noting the differences between this SCM and that laid out in \cite{Wang2025LatentGraphsFromSTData}.

The framework for spatiotemporal latent mechanism causal discovery we outline in the following sections provides researchers and decision makers ways to empirically derive causal models from complex spatiotemporal data and communicate those models accessibly.  

\section{Extending time series causal discovery to variables with spatial structure}\label{Extending}

\subsection{Time series causal discovery}\label{time series}

Temporal structures necessitate a treatment of autocorrelation and lagged effects, which can arise when values of a variable at one time are caused by values in the past \cite{Hong2009SerialDependence}. A variety of constraint-based algorithms have been proposed for time series causal discovery (see methods \cite{Assaad2021NoiseAndConstraintCD, Chen2026DiscreteTSCDLatentConf, Entner2010FCITimeSeries, Malinsky2018CDLatentConfTimeSeries, Ferdous2023NonStatAutoCorrTSCD, Moneta2006TimeSeriesParrCorr} and the review \cite{Gong2024TimeSeriesCDReview}). We focus on the popular PCMCI+ algorithm as its extension to accommodate latent temporal and dataset contexts is closest to meeting our goals in latent mechanism causal discovery \cite{Runge2020PCMCIplus, Gunther2023JointPCMCI}.

The PCMCI+ algorithm is an example of a constraint-based, PC-derived method to efficiently discover time-lagged and instantaneous relationships between nonspatial time series \cite{Runge2020PCMCIplus}. By shifting whole time series back by some time lag and including the shifted variables as other nodes in the system, it can explicitly test for autocorrelated relationships and time-lagged relationships between different variables that would be confounded or at least over-simplified in non-temporal causal discovery \cite{Runge2019PCMCI, Bystrova2024MultipleObsOneTimeCD}. 

PCMCI+ optimizes the PC algorithm for time series by splitting the task of causal discovery into two stages: one for lagged effects, and another for instantaneous. The lagged stage mirrors the PC algorithm, considering only edges between nodes in the ``present'' ($\mathbf{X}_t$) and ``past'' ($\mathbf{X}_{t-\tau}, \tau \in \{1, ..., \mathcal{T}\}$) where $\mathcal{T}$ is a hyperparameter indicated the maximum time lag considered. Since we assume effects cannot precede their causes, the directions of all edges discovered in the lagged phase are known and flow in the direction of time. It follows that no two present nodes can be causes of their shared neighbor in the past (i.e. the past neighbor cannot be a collider) as that would imply edges directed backwards in time. Furthermore, since only colliders can induce dependence, the inclusion by default of adjacent past nodes in separating sets $\mathbf{Z}_{it,jt} \subseteq \mathbf{X}\setminus\{X_{i,t}, X_{j,t}\}$ cannot falsely lead to failure to reject tests of $X_{i,t} \ind X_{j,t} \cond \mathbf{Z}_{it,jt}$. When searching for separating sets between present nodes in the instantaneous phase, we can therefore restrict the search space to sets including past nodes adjacent to $X_{i,t}$ and/or $X_{j,t}$ without worrying about potential colliders. Edge direction proceeds as for the PC algorithm \cite{Spirtes1991PC}. Most assumptions of the PC algorithm are therefore also made by PCMCI+; the time series algorithm assumes causal faithfulness, sufficiency, and stability of relationships across time \cite{Runge2020PCMCIplus}. 

\subsection{Causal discovery across contexts}

The joint causal inference framework introduced in \cite{Mooij2020JCI} has been applied to extend PCMCI+ to time series collected across different contexts and/or compiled from different datasets \cite{Gunther2023JointPCMCI}. Contexts are defined as sets of exogenous variables that may impact, but are not impacted by, system causal relationships. For example, the context of how cold it was over the winter will vary across ecosystems and likely cause changes to wildlife population sizes, but we would not expect changes in wildlife abundance to affect the weather. 

Joint PCMCI+ (J-PCMCI+) can efficiently discover the correct graphical representation of systems generated from a SCM involving time- and dataset-dependent system variables $\mathbf{X}_t^{(d)} = \{X_{i,t}^{(d)}\}_{i \in \mathcal{I}}$ and observed and latent context variables that vary by either time $t$ or, disjointly, by dataset $d$, $\mathbf{C} = \mathbf{C}_{\text{time}} \sqcup \mathbf{C}_{\text{dataset}}$, such that, for $i \in \mathcal{I}$:
\begin{equation}\label{eq:JPCMCI SCM}
    \begin{split}
        & \mathbf{X}_t^{(d)} := \mathbf{f}(\mathbf{pa}_{\mathbf{X}}(\mathbf{X}_t^{(d)}), \mathbf{pa}_{\mathbf{C}_{\text{time}}}(\mathbf{X}_t^{(d)}), \mathbf{pa}_{\mathbf{C}_{\text{dataset}}}(\mathbf{X}_t^{(d)}), \mathbf{\epsilon}_t^{(d)}) \\
        & \mathbf{C}_{\text{time},t} := \mathbf{g}(\mathbf{pa}_{\mathbf{C}_{\text{time}}}(\mathbf{C}_{\text{time},t}), \mathbf{\epsilon}_{\text{time},t}) \\
        & \mathbf{C}_{\text{dataset}}^{(d)} := \mathbf{h}(\mathbf{pa}_{\mathbf{C}_{\text{dataset}}}(\mathbf{C}_{\text{dataset}}^{(d)}), \mathbf{\epsilon}_{\text{dataset}}^{(d)})
    \end{split}  
\end{equation}
where the noise terms $\mathbf{\epsilon}_t^{(d)}, \mathbf{\epsilon}_{\text{time},t},$ and $\mathbf{\epsilon}_{\text{dataset}}^{(d)}$ are jointly independent and identically distributed across both time and datasets, just time, and just datasets, respectively. The sets of functions $\mathbf{f, g}$ and $\mathbf{h}$ define system variables $\mathbf{X}$, time-dependent contexts $\mathbf{C}_{\text{time},t}$ and dataset-dependent contexts $\mathbf{C}_{\text{dataset}}^{(d)}$, respectively. The set $\mathbf{pa}_{\mathbf{X}}(\mathbf{X}_t^{(d)})$ indicates those parents of $\mathbf{X}_t^{(d)}$ in $\mathbf{X}$, and analagously for $\mathbf{pa}_{\mathbf{C}_{\text{time}}}(\mathbf{X}_t^{(d)})$ and $ \mathbf{pa}_{\mathbf{C}_{\text{dataset}}}(\mathbf{X}_t^{(d)})$.

J-PCMCI+ proceeds by constructing ``dummy'' variables from embeddings of dataset and temporal indices to stand-in as proxies for latent contexts. In doing so, they assume that latent dataset and temporal contexts are deterministic functions of dataset and time indices, respectively \cite{Gunther2023JointPCMCI}. The algorithm then tests for links between observed contexts and system variables, between dummy variables and system variables, and finally between system variables. It returns a ``dummy-deleted'' graph showing only the links concerning observed contexts and system variables.

Note our slight departure of notation from \cite{Gunther2023JointPCMCI}; this is to ensure comparability across SCMs and to clarify context definitions. In the original formulation, dataset contexts are termed ``spatial'' contexts. We argue that this is a misnomer in that there is no geometry invoked in the structure of datasets, and that context-associated variance is defined to be iid across datasets \cite{Schabenberger2017SpatialStats}. Though geometry may be invoked in the construction of dummy variables that represent latent contexts, this definition does not allow for spatial autocorrelation and confounding. In fact, ``spatial'' contexts are simulated as iid draws from a standard normal distribution in numerical experiments \cite{Gunther2023JointPCMCI}.

Many similarities exist between the SCM we defined for latent mechanism causal discovery (Equation~\ref{eq:Latent Mech SCM}) and that of \cite{Gunther2023JointPCMCI} (Equation~\ref{eq:JPCMCI SCM}). Both assume exogeneity of contexts/confounders and stationarity of relationships between variables in $\mathbf{X}$. The principle difference between (\ref{eq:Latent Mech SCM}) and (\ref{eq:JPCMCI SCM}) is that we assume in (\ref{eq:Latent Mech SCM}) that all latent spatial variables $\mathbf{U}$ are parents of all system variables $\mathbf{X}$. The definitions of latent confounders also differ:  $\mathbf{U}$ are fixed, measurable functions of spatial structure $S$ in (\ref{eq:Latent Mech SCM}), whereas dummy contexts are deterministic functions of dataset indices in J-PCMCI+ \cite{Gunther2023JointPCMCI}.

The use of dummy contexts adds as many new variables as locations where data were collected. This significantly increases the complexity of causal discovery, and thus finite-sample multiple-testing error, without actually adding any spatial information. If we seek to discover the graph of a spatially-structured time series system as in Equation~\ref{eq:Latent Mech SCM}, however, we may be able to build from the logic of \cite{Gunther2023JointPCMCI}. Here we turn to the literature for causal effect estimation where explicit adjustments for spatial confounding have been relatively better explored.

\subsection{Latent spatial confounding in causal effect estimation}\label{Spatial confounding}

Spatial confounding of a causal effect occurs when bias is introduced by omission of a variable with spatial structure \cite{Gilbert2024CISpatConf}. Where spatial confounding is known to exist and cannot be controlled for by observed variables, spatial structure itself can be used as a proxy for even unquantifiable confounding \cite{Schnell2020SpatConfDisease}. 

Identifiability of such effects requires certain assumptions. \cite{Gilbert2024CISpatConf} show that the effect of exposure $X_i$ on outcome $X_j(x_i)$, where they are both confounded by unmeasured variables $\mathbf{U}$ such that $\mathbf{U}$ are fixed, measurable functions of spatial structure $S$, is identifiable given spatial structure is not a collider with regards to $X_i$ and $X_j(x_i)$ and given positivity of the exposure with respect to $S$. The same results hold conditional on any measured confounders, if necessary. Unbiased causal effects can thus be estimated by controlling for spatial structure, given a sufficiently flexible nonparametric model is used. A discussion grounding this treatment of spatial confounders in the instrumental variables literature can be found in \cite{Woodward2025IVSpatConf}.

This method is highly sensitive to the spatial scale of unobserved confounders relative to that of observed variables \cite{Paciorek2010ScaleSpatConf}. Specifically, the resolution of unobserved confounders must be coarser than that of observed variables; effects are unidentifiable when there are strata of unobserved confounders for which only one value of an observed variable is possible \cite{Schnell2020SpatConfDisease}. Given that unobserved confounders are, obviously, unobserved, this is an untestable assumption and must be made based on domain knowledge and taken into account in interpretation.

\section{A framework for latent mechanism causal discovery}

Identifiability of effects between variables defined by the latent mechanism SCM in Equation~\ref{eq:Latent Mech SCM} is guaranteed by this assumption set \cite{Gilbert2024CISpatConf}. Spatial confounders $\mathbf{U}$ are defined identically, and spatial structure $S$ cannot be a collider by its exogenous definition. 

We therefore propose an alteration to PCMCI+ in which all tests of independence condition on spatial structure. We show that estimation of the graph $\widehat{\mathcal{G}}$ over observed variables $\mathbf{X}$ by such an algorithm is sound and complete.

\begin{theorem}\label{thm:Discovery}
    Let the system $\mathcal{X}$ be a sufficient SCM of the form described in Equation~\ref{eq:Latent Mech SCM}. Suppose there is an acyclic time series graph $\mathcal{G}$  with regards to which the joint distribution $\mathbb{P}(\mathbf{X}, \mathbf{U})$ induced by the SCM is both faithful and satisfies the causal Markov condition. Then, given positivity of variables $X_i \in \mathbf{X}$ with respect to $S$, and assuming consistency of conditional independence tests, the estimated graph $\widehat{\mathcal{G}}$ corresponds to $\mathcal{G}$ in the large size of $S$ and length of time series limit.
\end{theorem}

Proof of this theorem relies on the fact that although $S$ and $\mathbf{U}$ may \textit{confound} relationships between variables in $\mathbf{X}$, they may not mediate or be a collider by their exogenous definitions \cite{Gunther2023JointPCMCI}. Therefore, any two variables $X_i, X_j; i \neq j$ that are not connected by an edge in $\mathcal{G}$ and separable given the set $\mathbf{Z} \subset \mathbf{X}$ are independent conditional on $(\mathbf{Z}, S, \mathbf{U})$ by the causal Markov condition (formally termed \textit{d-separation}, see \cite{Pearl2009Causality} for further definition). The result extends to conditioning on just $(\mathbf{Z}, S)$, since $S$ is a measurable function of the double $(S, \mathbf{U})$, and therefore invokes the same conditional distribution \cite{Gilbert2024CISpatConf}. Therefore, given faithfulness, tests of $X_i \ind X_j \cond \mathbf{Z}, S$ correctly lead to deletion of edges in $\widehat{\mathcal{G}}$ that are not in $\mathcal{G}$. This establishes soundness of the algorithm. Completeness follows from soundness given completeness of the edge orientation phase of PCMCI+ \cite{Runge2020PCMCIplus, Gilbert2024CISpatConf}. A full proof is available in supplementary material \cite{Supple2026Supplement}.

In the following sections, we explore the assumptions, limitations, practicalities, and potential of latent mechanism causal discovery for spatially-structured time series. 

\subsection{Conditional independence testing}\label{GCM}

The rigor of constraint-based causal discovery hinges on consistent, accurate conditional independence testing in addition to its assumptions. Algorithms tend to be independence test-agnostic, however, and rarely enforce the use of any particular test or make claims of asymptotic correctness beyond the oracle setting where independence tests make no errors \cite{Runge2019PCMCI, Peters2013TiMINo, Glymour2019CDReview, Cinelli2025Challenges}. In fact, our proof of Theorem~\ref{thm:Discovery} is a little cavalier regarding assumed consistency of conditional independence tests.

This assumption is difficult to meet. \cite{Shah2020GCM} proved that there is no nontrivial test that can distinguish between the null and alternative hypotheses of conditional independence and dependence with consistent power greater than the significance level. Restrictions must be placed on the null to ensure power at any alternative.

\cite{Shah2020GCM} propose the generalized covariance measure (GCM), as a practical approach to testing conditional independence. The null hypothesis of this test is that $X_i$ and $X_j$ are independent given $\mathbf{Z}$:
\begin{equation*}
    \begin{split}
        \text{H}_0 : \, & X_i \ind X_j \cond \mathbf{Z}, \, \text{or, equivalently, } \\
        & \, \mathbb{P}(X_i, X_j \cond \mathbf{Z}) = \mathbb{P}(X_i \cond \mathbf{Z}) \mathbb{P}(X_j \cond \mathbf{Z}).
    \end{split}
\end{equation*}

The GCM test approximates the conditional distributions $\mathbb{P}(X_i \cond \mathbf{Z})$ and $\mathbb{P}(X_j \cond \mathbf{Z})$ with functions $\widehat{f}$ and $\widehat{g}$ estimated by regressing $X_i$ on $\mathbf{Z}$ and $X_j$ on $\mathbf{Z}$, respectively. A normalized sum similar to partial correlation of the product of residuals from fitted $\widehat{f}$ and $\widehat{g}$ is calculated. If the absolute value of this statistic is significantly large, the null hypothesis of independence is rejected. Consistency of the test is guaranteed given the product of mean squared error of $\widehat{f}$ and $\widehat{g}$ is on the order of $n^{-1}$ and that the products of mean squared error of $\widehat{f}$ with the noise associated with $X_j$, and mean squared error of $\widehat{g}$ with the noise associated with $X_i$, is on the order of $1$ \cite{Shah2020GCM}. See supplementary material \cite{Supple2026Supplement} for further technical details and Section~\ref{Uncertainty} for a discussion of significance thresholds.

No particular function or model class is proposed to be ideal; \cite{Shah2020GCM} use kernel ridge regression in examples, while \cite{Scheidegger2022WGCM} use generalized additive models (GAMs) and boosted regression trees in their extension of the GCM test to nonlinear dependence. \cite{Gilbert2024CISpatConf} test a number of flexible, non-parametric models for their ability to recover spatially-confounded causal effects and find best performance is achieved by GAMs with three dimensional splines over location and exposure, Bayesian additive regression trees (BARTs), and versions of the GAMs and BARTs fit via double machine learning. 

Given the iterative nature of constraint-based causal discovery, we would like to find a model to approximate $\widehat{f}$ and $\widehat{g}$ that is computationally efficient, while also meeting conditions for spatial de-confounding and the consistency of the GCM test. We suggest that the use of GAMs with three dimensional splines over each conditional variable $Z_r \in \mathbf{Z}$ and spatial coordinates $S = \{S_1, S_2\}$ fit these criteria. Specifically, estimation of the function $f$ via the equation:
\begin{equation}
    \begin{split}
    & \widehat{f}^{(n)} = \arg \min_{f} \frac{1}{n} \sum_{k=1}^{n} \left[ \left(X_{i,k} - f(\mathbf{Z}_k, S_{1k}, S_{2k}) \right)^2 \right]  \\ 
     & + \lambda \sum_{r = 1}^{|\mathbf{Z}|}J_{m = 2}^{D = 3}(f)_r 
    \end{split}
\end{equation}
where $\lambda \geq 0$ controls the strength of the three-dimensional ($D = 3$), second-order ($m = 2$) thin plate spline smooth penalty term $J_{m = 2}^{D = 3}$, meets assumptions due to its properties as a reproducing kernel Hilbert space and given finite higher-order moment conditions \cite{Wood2017Smoothers, Wahba1990SplineTheory}. See further definitions and details in the supplementary material \cite{Supple2026Supplement}.

We note that there is a wide literature on conditional independence testing, with methods falling broadly into kernel-based (\cite{Fukumizu2007KernelCondDep, Song2022KernelIndTestCluster, Gretton2005HSIC}), regression-based (GCM \cite{Shah2020GCM}; see also \cite{Scheidegger2022WGCM, Christgau2023NonParCondIndTest, Strobl2026FastRegCondInd, Zhang2022ResidualCondInd}), and mutual information-based schools (\cite{Runge2018MutualInfo}). We focus here on regression-based methods, and especially GCM, for their theoretical overlap with the model selection principles of effect identification under spatial confounding, but do not claim optimality \cite{Gilbert2024CISpatConf}.

\subsection{Assumptions}\label{Assumptions}

If we accept consistency of conditional independence tests, we still have to address the reasonableness of causal discovery and spatial assumptions. Exogeneity of spatial structure is necessary, but it is generally met given a standardized or randomized sampling design was used \cite{Wang2012SpatSample}. If locations were sampled because they were expected to exhibit certain trends, however, the assumption will not hold. Practitioners should assess the sampling design(s) under which their data were collected, and consider methods to reduce dependence of locations on data values when necessary. For example, citizen science datasets in which volunteers submit wildlife observations are well known to have spatial bias towards areas accessible to people and/or suitable for outdoor recreation \cite{Backstrom2024CitSciBias}. Down-sampling to ensure relatively even spatial coverage may alleviate this issue, and could even provide opportunities to investigate algorithm variance if different sub-samples are used for a bootstrap-like procedure (see Section~\ref{Uncertainty} for further discussion).

\subsubsection{Faithfulness}\label{faithfulness}

Given noise, measurement error, and the potential violation of untestable assumptions, practitioners may question if conditional independence relationships as discovered from data faithfully correspond to true causal structures. This is exacerbated by algorithm design. To improve computational and statistical efficiency, PC-based algorithms search for the \textit{minimum} separating set for each pair of variables, rather than for \textit{all} separating sets. That is, the exclusion of a variable $X_a$ from a separating set $\mathbf{Z}_{ij}$ found in a PC-based algorithm does not in all cases indicate that $X_i \notind X_j \cond (\mathbf{Z}_{ij}, X_a)$, but just that a smaller set was found first \cite{Ramsey2006Faithfulness}.

Options to address design-based faithfulness violations are already included in PCMCI+ \cite{Runge2020PCMCIplus}. The algorithm allows users to specify how strictly they want to determine if  $X_a$ is a collider in the structure $X_i \edge X_a \edge X_j$; practitioners may choose to be more conservative with regards to faithfulness at the cost of higher complexity.

\subsubsection{Sufficiency and unobserved non-spatial confounders}

As discussed in \cite{Gilbert2024CISpatConf}, only those unobserved confounders with spatial structure will be accounted for by the inclusion of spatial contexts. Unobserved variables \textit{without} spatial structure may still confound discovered relationships, leading algorithms to misdirect edges.

Many constraint-based algorithms have been developed to allow for latent confounders (i.e.~relax the causal sufficiency assumption). Often, extensions stem from the Fast Causal Inference (FCI) algorithm proposed by \cite{Spirtes1999FCI}, which, in concert with the additional rules proposed in \cite{Zhang2008FCI}, is provably sound and complete under sufficiency violations. In FCI-related algorithms, edges are not directed but instead marked at each end with either a tail, --, to indicate ancestry or an arrowhead, >, to indicate descendants. This means that double-headed arrows, $\leftrightarrow$, may be discovered, indicating latent confounding. Where the algorithm cannot decide, it leaves a circle marking $\circ$ \cite{Spirtes1999FCI}. FCI has been extended to time series settings in \cite{Malinsky2018CDLatentConfTimeSeries} and \cite{Entner2010FCITimeSeries}. An alternative pathway generalizes the momentary conditional independence testing central to PCMCI+ and its extensions to insufficient settings \cite{Gerhardus2020LatentPCMCI, Rabel2025LatentContext}. 

The conditional independence testing phase of FCI-like algorithms is usually identical to that of PC-like algorithms; they differ in edge orientation rules only \cite{Spirtes1999FCI, Zhang2008FCI}. Our spatial extension alters only the conditional independence testing phase of a constraint-based algorithm since we assume exogeneity of spatial structure and thus direction of its effect is known. If practitioners are not willing to make assumptions of nonspatial sufficiency, they could hypothetically swap out sufficiency-assuming rules for edge orientation for those of insufficient algorithms. Those willing to assume there are no latent nonspatial confounders can enjoy the relatively higher power that comes with causal sufficiency \cite{Pearl2009Causality}. An insufficient setting extension along these principles is an area of active research.

\subsubsection{Stability and temporal stationarity}

We make the implicit assumption that relationships between variables do not change over time. This may be relaxed. J-PCMCI+ allows for temporal contexts that may capture some of that nonstationarity \cite{Gunther2023JointPCMCI}, and there is interesting work on breaking up time series into “regimes” under which different relationships between variables may be at play \cite{Saggioro2020RegimeCD, Mameche2025NonStatTSCD, Rabel2025LatentContext}. Work would be needed, however, to incorporate that structure into causal discovery algorithms that are also spatially explicit and to determine how large a sample size is needed to achieve power with that much flexibility.

\section{Perspectives and future directions}

In this section, we discuss remaining challenges and open questions in the development of a causal discovery algorithm for spatially-structured time series. We then outline specific potential applications and general intended impact of our work.

\subsection{Algorithm implementation and evaluation}

\subsubsection{Uncertainty}\label{Uncertainty}

Most algorithms do not provide any measure of uncertainty in edge assignment and/or direction. Bootstrapping is a natural choice when parametric estimates of variance seem impossible, but requires a large dataset whose structure would not necessarily be disrupted by resampling. \cite{Debeire2024BootstrapPCMCI} propose a bootstrap aggregation procedure for the PCMCI+ algorithm that avoids disrupting time series structures by resampling from the shifted data. Spatially stratified resampling (such as in the downsampling example in Section~\ref{Assumptions}) may allow this technique to be extended to latent mechanism causal discovery. Early tests, however, do not demonstrate the same power improvement as shown by \cite{Debeire2024BootstrapPCMCI}, and, depending on dimensionality, rerunning an already complex algorithm enough times to generate a reasonable sample may be prohibitive.

\cite{Petersen2021LongitudinalCD} alternatively propose reiterating a causal discovery algorithm across many significance thresholds to capture the ``strength of support'' for, or certainty of, each causal relationship. More certain relationships should continue to appear in graphs with stricter thresholds that admit fewer edges. Complementarily, we propose that uncertainty could be communicated in a single graph by reporting an aggregate across thresholds with edges labeled by their frequency, weighted by the relative magnitude of each significance threshold.

\subsubsection{Computational intensity}

One of the largest challenges in developing causal discovery algorithms is computational intensity \cite{Le2019ParallelPC}. Despite choosing models and designing the algorithm to be as time-efficient as possible without losing power, the sheer task of fitting all models to approximate conditional distributions for all pairs of variables conditional on iterative sets of the rest of the system is gargantuan \cite{Pedersen2019HGAMs}. 

Parallelization offers a promising path to faster overall algorithms, but is difficult in practice due to the interdependence of stages of causal discovery (i.e.~which conditional independence tests are run depends on findings from previous testing). Balancing the efficiencies gained from parallelization and from information sharing across conditional independence tests, \cite{Le2019ParallelPC} propose parallel-PC, which searches for separating sets for different variable pairs across multiple cores and then synchronizes information at each iteration. These adjustments significantly reduce runtime compared to the order-independent PC algorithm. Efficiency improved with more cores even on very large datasets. 

Algorithm design choices made to reduce computational complexity, however, are often at odds with the notion of dividing distinct tasks across multiple cores. For example, approximating models for the GCM test can be more efficiently fit by utilizing parallelization features already built into computational packages such as \verb|mgcv| \cite{Wood2015BAM}. Tests of a group of separating sets can be prioritized by the test statistic associated with previous tests of similar sets, increasing the likelihood that you find the separating set sooner and need to perform fewer tests overall \cite{Runge2019PCMCI}. The residuals of approximating models can be saved and accessed from temporary storage, eliminating the need to re-fit models for each new GCM test. All of these features require information sharing to inform each test and/or access to multiple cores, neither of which play nicely with parallelization. Further testing will inform whether efficiencies lost in parallelization outweigh efficiencies gained in the spatiotemporal context.

\subsubsection{Simulations and sensitivity analysis}

All causal discovery algorithms rely on untestable assumptions. Where real-world truth is unknowable, simulation studies help us benchmark the performance and robustness of an algorithm to those assumptions \cite{Cinelli2025Challenges}. Although spatially-structured time series abound in real-world data (e.g.~abundance of bird species over time in Europe \cite{Rigal2023BirdCCM}, Dengue transmission across time in the Philippines \cite{Sanchez2025CausalEpi}, and air quality monitoring near dump sites in Indonesia \cite{Considine2026PlasticEffectSTCI}), their simulation is not so straightforward \cite{Glake2020SpatTempABM}. This is further complicated by a desire to test the ability of an algorithm to recover potentially nonlinear, non-Gaussian relationships. 

Most data for tests of causal discovery algorithms are simulated from additive noise hierarchical models in which each variable is a function of its direct cause(s). The additive noise at each level of the model compounds down the causal order such that descendant variables have larger marginal variances than their ancestors. Causal discovery algorithms can then take advantage of the predictable variability structure to identify graphs more successfully than they could for standardized or randomly varying systems \cite{Reisach2021BewareSimDAG}. This issue is more relevant for methods that exploit asymmetries in noise than constraint based methods, but standardization and/or careful model design are needed to avoid inflating discovery performance on all simulated data sets.

Agent-based models (ABMs, also known as individual-based models in some fields) may provide a more appropriate method of simulating data to benchmark and check sensitivity of causal discovery algorithms. In ABMs, complex system properties arise from mechanistic rules dictating individual agents' interactions with their environment, a philosophy that nicely mirrors the goal of latent mechanism causal discovery \cite{Grimm2005IBMEcology}. Guides for ABM design suggest visualizing the models with influence diagrams \cite{Grimm2005IBMEcology} and an emerging literature calls for use of causal discovery in ABM validation \cite{Janssen2022CDtoDesignABM, Yu2024CDforABMValidation}. Because noise arises from the randomness of agents' behavior rather than by sequential addition, data simulated by ABMs are not subject to the marginal variance issues described for hierarchical models.

ABMs provide a convenient simulation framework to test violations of an algorithm's assumptions. One could alter, for example, the spatial scale of the environment or the temporal scale at which agents act to check for algorithm robustness to scale mismatch. Variables relevant to agents' actions in the simulation can be left out to assess algorithm performance in light of unobserved confounders. More work is needed, however, to evaluate if the potential of ABMs for robust causal discovery benchmarking is worth the consideration and computational power needed to simulate them \cite{Grimm2005IBMEcology}.

\subsubsection{Causal discovery performance metrics}

The literature on evaluating the performance of causal discovery algorithms is sparse relative to that for building them. CauseMe, an online database for sharing and comparing causal discovery algorithms \citep{Runge2019TSCDReview}, uses machine learning classifier metrics such as true positive rate, false positive rate, F1, and area under a receiving operator curve to compare performance of a wide variety of algorithms on benchmark datasets. While reasonable at face value, \cite{Petersen2025NegativeControls} has shown that the expected F1 score of a random graph increases monotonically with the number of estimated edges; comparing graphs with different numbers of edges via F1 score (or other scores based on recall and precision) will bias towards more connected graphs. Instead, \cite{Petersen2025NegativeControls} suggest comparing graphs produced by an algorithm to a distribution of graphs created via random assignment of the same number of edges and propose an associated hypothesis test for differences between estimated graphs and random edge placement.

Other approaches describe metrics for testing the distance between graphs produced by an algorithm and the ground truth. Structural intervention distance describes the similarity of inferences one would make from estimated vs.~true graphs \citep{Peters2015StrIntDist}. It counts the number of changes one would have to make to an estimated graph to ensure that the structural causal model built from it matched that of the true graph. A recent extension to structural intervention distance allows for evaluation of partial graphs that admit unobserved confounders \citep{Henckel2024AdjIDDist}.

None of these methods are immediately applicable to the kinds of graphs produced by spatiotemporal causal discovery. The discussed options weigh all differences from the true graph evenly, but when considering the interpretation of time series causal discovery, some relationships are arguably ``more wrong'' than others. For example, in the system illustrated in Figure~\ref{fig:Illustrations}B, it would be ``more wrong'' relative to interpretation if an estimated graph had no connection between forest cover and species abundance than if it had estimated that as an instantaneous relationship. Qualitative comparisons that ask how and where interpretation or conclusion-making has gone wrong may be more practically useful. These questions can be asked in comparisons with both true graphs (for simulations) and/or a distribution of graphs produced by random edge placement \cite{Petersen2025NegativeControls}, and seem for now the best option to evaluate performance of spatiotemporal causal discovery algorithms. As discussed in Section~\ref{Uncertainty}, reporting weighted aggregates of graphs evaluated at different significance thresholds may provide information about consistency of estimates and contribute to evaluation.

\subsection{Applications and impact}

Studies that sought to use existing methods to identify underlying mechanisms of cause and effect from observed time series with spatial structure have yet to achieve the goals of latent mechanism causal discovery. Those who employed iid time series algorithms such as PCMCI+ to analyze spatially structured data reported poor performance and lack of interpretability \cite{Nichol2023PCMCIBenchmarkST, Miersch2025PCMCIFlood}. Other studies used spatial convergent cross mapping, a method that discovers pair-wise, quasi-causal links between independently replicated time series \cite{Rigal2023BirdCCM, Sarfo2024EffectRuralLandscape, Clark2015SpatialCCM, Barraquand2020SpeciesInteractionsCCM}. These studies produced evidence of potential causal relationships between pairs of variables, but were not able to combine these relationships to holistically understand the larger system. Neither approach addressed spatial confounding or took advantage of the known spatial structure of their data. The framework we outline, though not yet stress-tested or benchmarked, takes an important first step towards achieving causal discovery of mechanisms from spatiotemporal data.

We expect work in this area to have impact outside of statistical theory, as visualizations allow non-specialists to engage in analysis, interpretation, and decision making. In ecology and environmental sciences, causal graphs interface nicely with existing visual policymaking methods. The United States Environmental Protection Agency, for example, advocate for the use of qualitative influence diagrams (QIDs) in describing complex systems for environmental policymaking \cite{Carriger2018QIDs}. QIDs visually map controlling factors and impacts to help policymakers decide on interventions and evaluate downstream effects. Methods exist for deriving conditional distribution information from QIDs, and empirical causal graphs can easily be read as influence diagrams \cite{Wellman1990QualGraphs}. Further examples of graphical methods in sustainable development can be seen in \cite{Videira2013}. 

In public health and epidemiology, life-course epidemiology models that explore how various risk factors affect health outcomes across a lifetime are often represented as graphs. \cite{Petersen2021LongitudinalCD} compared empirical discovery of causal relationships between socioeconomic factors and the development of depression to expert-derived graphs. Spatiotemporal latent mechanism causal discovery could strengthen these results by accounting for the spatial non-independence of studied individuals. \cite{Raunig2023EconGraph} conducted a country-by-country graph-based causal analysis of the relationships between economic policy uncertainty and stock market volatility that would similarly benefit from a spatially-explicit approach.

\subsection{Future directions and conclusion}

We are currently working to implement this framework and test both its performance on simulated data and its utility for practical applications. Specifically, we are investigating sensitivity to violations of the scale condition (Section~\ref{Assumptions}) and extensions to insufficient settings and discrete space structures. 

A good causal understanding of the world is key to effective decision making. The extension of causal discovery algorithms to time series with spatial structures described here provides a new path to causal understanding of the complex spatiotemporal structures of data in many critical fields such as public health, economics, environmental sciences, and ecology. 

\begin{acks}[Acknowledgments]
We thank Dave Miller for his insights on spatial generalized additive modelling in the context of causal discovery.

We are grateful to anonymous referees for their helpful comments and revisions.
\end{acks}

\begin{funding}
R.F.S. was supported by a studentship from the UKRI EPSRC (project 2921649). B.S. would like to thank the Isaac Newton Institute for Mathematical Sciences, Cambridge, for support and hospitality during an INI retreat (EPSRC grant EP/Z000580/1), where some work on this paper was undertaken. 
\end{funding}

\begin{supplement}
\stitle{Supplement to ``Discovering causal relationships
between time series with spatial structure''}
\sdescription{Proof of Theorem 4.1, details of the Generalised Covariance Measure (GCM) test, and demonstration of the utility of a spatial generalized additive model for GCM testing. Available \href{https://becca-supple.github.io/files/papers/Supple_Supplement.pdf}{here}.}
\end{supplement}

\bibliographystyle{imsart-number} 
\bibliography{references}       

\end{document}